\documentclass[aps,prl,twocolumn,showpacs]{revtex4}
\usepackage{graphicx}
\usepackage{dcolumn}
\usepackage[colorlinks=true,dvipdfm]{hyperref}

\begin{document}

\title{Experimental simulation of fractional statistics of abelian anyons in
the Kitaev lattice-spin model}
\author{Jiang-Feng Du$^{1}$}
\email{djf@ustc.edu.cn}
\author{Jing Zhu$^1$}
\author{Ming-Guang Hu$^{2}$}
\author{Jing-Ling Chen$^{2}$}
\affiliation{$^{1}$Hefei National Laboratory for Physical Sciences at Microscale and
Department of Modern Physics, University of Science and Technology of China,
Hefei, Anhui 230026, People's Republic of China\\
$^{2}$Theoretical Physics Division, Chern Institute of Mathematics, Nankai
University, Tianjin 300071, China}
\date{\today}

\begin{abstract}
In two-dimensions, the laws of physics even permit the existence of anyons
which exhibit fractional statistics ranging continuously from bosonic to
fermionic behaviour. They have been responsible for the fractional quantum
Hall effect and proposed as candidates for naturally fault-tolerant quantum
computation. Despite these remarkable properties, the fractional statistics
of anyons has never been observed in nature directly. Here we report the
demonstration of fractional statistics of anyons by simulation of the first
Kitaev lattice-spin model on a nuclear magnetic resonance system. We encode
four-body interactions of the lattice-spin model into two-body interactions
of an Ising spin chain in molecules. It can thus efficiently prepare and
operate the ground state and excitations of the model Hamiltonian. This
quantum system with convenience of manipulation and detection of abelian
anyons reveals anyonic statistical properties distinctly. Our experiment
with interacted Hamiltonian could also prove useful in the long run to the
control and application of anyons.
\end{abstract}

 \maketitle

Particles in nature behave at two distinct statistics referring to bosons
and fermions. But if one is restricted to two-dimensional (2D) systems, a
class of fractional statistics intervening between bosonic and fermionic
statistics appears. It is determined by some quasi-particles known as anyons%
\cite{1982Wilczek,1982Wilczek2}. The quantum state of anyons can acquire an
unusual phase when one anyon is exchanged with another one, in contrast to
usual values +1 for bosons and -1 for fermions. In general, the quantum
state of the $n$ indistinguishable particles belongs to a Hilbert space that
transforms as a unitary representation of the braid group. Abelian anyons
are defined to be these particles that transform as one-dimensional
representation of the braid group, while the nonabelian anyons correspond to
representations that are of dimension greater than one \cite{2004Preskill}.

Anyons have intrigued great interests not only for their unusual fractional
statistics but rather for reason of their crucial role in topological
quantum computation (TQC) \cite%
{2003Kitaev,2006Kitaev,1997Preskill,2003Mochon,2004Mochon,2000Freedman,2002Freedman}%
, since exotic exchange statistics on 2D plane endows anyons with nontrivial
topological properties that indicates their potential values in use.
Although anyons have been responsible for the fractional quantum Hall system
\cite{1982Tsui,1983Laughlin}, a direct observation of fractional statistics
associated with anyon braiding is hard in this system and has attracted
several intriguing theoretical proposals \cite%
{2005Sarma,2006Bonderson,2006Stern,2007Weeks} etc.. Particularly, the
exactly solved models for naturally fault-tolerant topological quantum
computation proposed by Kitaev \cite{2003Kitaev,2006Kitaev} demonstrate the
excitations with anyonic features and thus also provide a direction for
experimental detection of anyons \cite{2003Duan}.

Recently, trapped ions, photons, or nuclear magnetic resonance (NMR) were
suggested to realize a small-scale system for proof-of-principle
demonstration of the anyon braiding statistics based on the first Kitaev
lattice-spin model \cite{2007Han}. Experimental demonstrations by using
photons have been made with four \cite{2007Pachos} or six \cite{2007Lu}
photons. Yet since the background Hamiltonian vanishes in such optical
systems, the defect that they are not protected from noise and the particle
interpretation of the excitations is ambiguous was soon pointed out \cite%
{2007Jiang}.

In this letter, inspired by the four-photon experiment \cite{2007Pachos}, we
start by simulation of the first Kitaev lattice-spin model with the
interacted nuclear spin-1/2 particles in molecules to demonstrate the
fractional statistics of anyons. A significant advantage of using nuclear
spin system is that the concept of anyons is originally and generally
considered with the interacted spin particles in magnetic field and there
are well established nuclear magnetic techniques for coherent control and
measurement of nuclear spin systems \cite{2004Vandersypen}. This allows us
to prepare the initial multi-spin entangled state of spins, directly
simulate dynamic evolution of anyonic quasi-particles with the Kitaev
lattice-spin model Hamiltonian, and detect the abelian anyons.

\begin{figure*}[tbp]
\centering
\includegraphics[width=1.8\columnwidth]{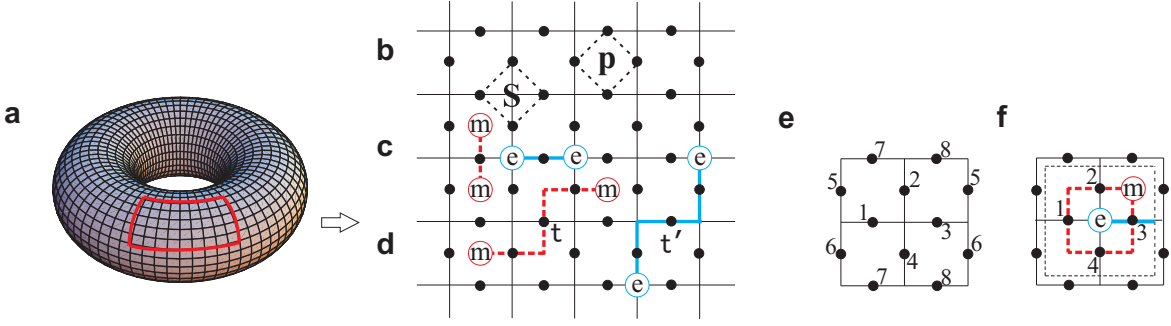}\newline
\caption{(Color online) An illustration of the first Kitaev model: (a) A $%
k\times k$ square lattice residues on a torus; (b) The hermitian operator $%
A_s$ ($B_p$) is defined on a star $s$ (face $p$) and acts on the four qubits
of the star (face); (c) Two $m$ particles are created as $|m\rangle=\protect%
\sigma_r^x|\protect\xi\rangle$ at the both faces around the $r$-th qubit and
two $e$ particles are created as $|e\rangle=\protect\sigma_r^z|\protect\xi%
\rangle$ at the both vertices around the $r$-th qubit; (d) The string
operator $S(t)$ acting on the ground state creates two $e$ ($m$) particles
around the first acted qubit and then move one particle along the path $t$ ($%
t^{\prime}$); (e) The minimal system is a $2\times2$ square lattice with
eight qubits on edges; (f) A braiding process can be achieved with only four
qubits.}
\label{fig1}
\end{figure*}

First let us review the essence of the first Kitaev lattice-spin model \cite%
{2003Kitaev}. Considering a $k\times k$ square lattice on a torus (see Fig. %
\ref{fig1} (a)), one spin or qubit is attached to each edge of the lattice.
Thus there are $2k^2$ qubits. For each vertex $s$ and each face $p$,
consider operators of the following form:
\[
A_s=\prod_{j\in \text{star}(s)}\sigma_j^x,\quad B_p=\prod_{j\in \text{%
boundary}(p)}\sigma_j^z,
\]
where the $\sigma_j^\alpha$ denotes the Pauli matrix ($\alpha=x,y,z$) and it
acts on the $j$-th qubit of a star $s$ or boundary $p$ (see Fig. \ref{fig1}
(b)). These operators commute with each other because star $s$ and boundary $%
p$ have either 0 or 2 common qubits. The operators $A_s$ and $B_p$ represent
four-body interactions and they are Hermitian with eigenvalues $1$ and $-1$.
Putting them together, it constructs the model Hamiltonian
\[
H=-\sum_{s\in \text{torus}}A_s-\sum_{p\in \text{torus}}B_p .  \label{eq-Hami}
\]
Its four-fold degenerate ground states $\{|\xi\rangle\}$ satisfy $%
A_s|\xi\rangle=|\xi\rangle$, $B_p|\xi\rangle=|\xi\rangle$ for all $s$, $p$
and construct a protected subspace. Hence it can define a quantum code
called a \emph{toric code}. Operators $A_s$ and $B_p$ are thus called the
\emph{stabilizer operators} of this code.

Here we concern more about anyons as the excitations of this Hamiltonian. A
pair of `electric charges' living on vertices (or `magnetic vortices' living
on faces) will be created at the both sides of the $r$-th qubit by the
operator $\sigma_r^z$ (or $\sigma_r^x$) acting on the ground state $%
|\xi\rangle$ (see Fig. \ref{fig1} (c)). The two kinds of particles are
denoted by $e$ and $m$ respectively. For the fusion rule ($e\times e=m\times
m=1$), one thus can define the so-called string operators
\begin{equation}  \label{eq-string}
S^z(t)=\prod_{r\in t}\sigma_r^z,\quad S^x(t^{\prime})=\prod_{r\in
t^{\prime}}\sigma_r^x,
\end{equation}
where $r\in t(^{\prime})$ means the $r$-th qubit on the path $t(^{\prime})$.
The operation $S^z|\xi\rangle$ (or $S^x|\xi\rangle$) first creates two $e$
(or $m$) particles at the two sides of the primarily acted qubit and then
moves one of them to the end of the path $t$ (or $t^{\prime}$) (see Fig. \ref%
{fig1} (d)). If one utilizes the string operators to move a $m$ (or $e$)
particle around an $e$ (or $m$) particle, we see that the total wavefunction
acquires a global phase factor $-1$. This is the unusual statistical
property of abelian anyons.

From the Fig. \ref{fig1} (a) and (e), we can see that the minimal system of
this Kitaev model is a $2\times2$ square lattice with eight independent
qubits on edges. The braiding statistics can be performed on this plane with
a $m$ particle winding one cycle around the other $e$ particle. Such a
series of operations only are related to, for example, the 1-st, 2-nd, 3-rd,
4-th qubits. So we can reduce the system to include merely these necessary
qubits and it will not compromise the physical meaning. In our experimental
setup, we would use such four-qubit system which could be viewed as a small
piece from a large one (Fig. \ref{fig1} (f)).

The Hamiltonian for this four-qubit system is $H=-\sigma_1^x\sigma_2^x%
\sigma_3^x\sigma_4^x-\sigma_1^z\sigma_2^z-\sigma_2^z\sigma_3^z-\sigma_1^z%
\sigma_4^z-\sigma_3^z\sigma_4^z$  with its ground state  $|\xi\rangle=(1/%
\sqrt{2})(1+\sigma_1^x\sigma_2^x\sigma_3^x\sigma_4^x)|0000\rangle$  that is
just GHZ state  $|\text{GHZ}\rangle=(|0000\rangle+|1111\rangle)/\sqrt{2}$.
In  principle, the creating operation of anyons can be performed on any one
of the four qubits. Here we choose the 3-rd qubit for the $e$ particle
creation, and the 2-nd, 1-st, 4-th, 3-rd qubits in order for $m$ particle
creation and transportation. In order to detect the global phase factor in
front of the total wavefunction after braiding, one needs to superpose the
initial state and the excitation state as $(|\xi\rangle+i|e\rangle)/\sqrt{2}$%
. Then after braiding the state becomes $(|\xi\rangle-i|e\rangle)/\sqrt{2}$.
It has the effect of transforming the unobservable global phase factor into
a visible local phase factor. By measuring the relative phase factor before
and after the braiding process, one can assure the fractional statistics of
abelian anyons.

\begin{figure}[tbph]
\centering
\includegraphics[width=1\columnwidth]{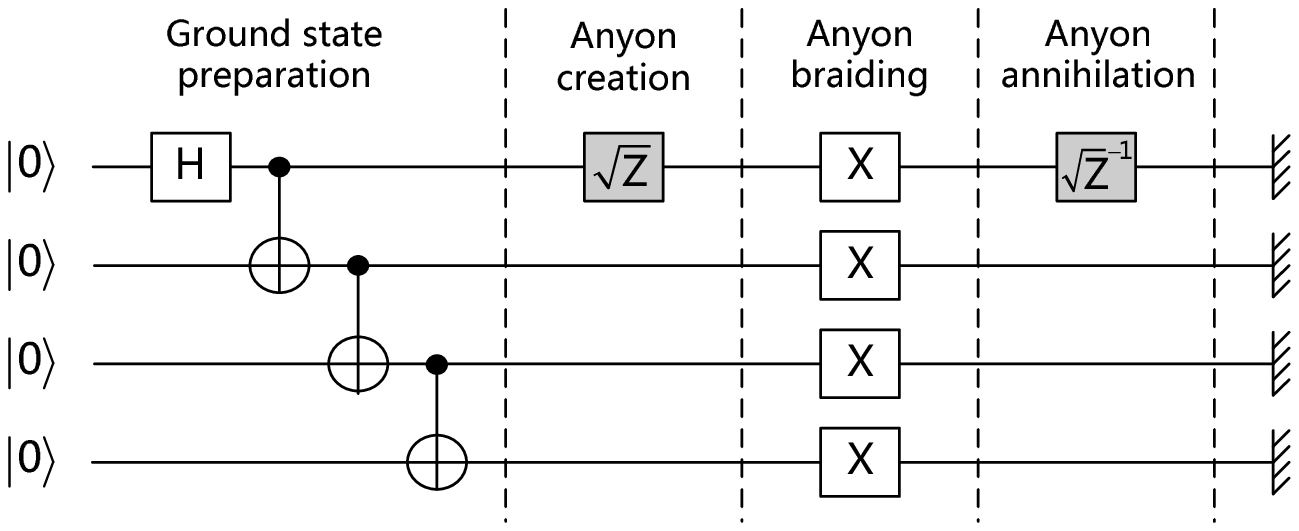}
\caption{The quantum network for the experiment. $H$ means a Hadamard
operation and $Z/X$ corresponds to $\protect\sigma_z$/$\protect\sigma_x$
operation. For the $\protect\sqrt{Z}$ operation, it has the function of
creating anyons and superposing the ground state and the anyonic excitation,
while the $\protect\sqrt{Z}^{-1}$ is its inverse operation.}
\label{fig2}
\end{figure}

The experimental network was shown as Fig. \ref{fig2}. In the experiment, we
first prepare a four-qubit GHZ state $|\text{GHZ}\rangle=(|0000\rangle+|1111%
\rangle)/\sqrt{2}$; then implement a $\sqrt{Z}$ operation that changes the
initial ground state to a superposed state $(|\xi\rangle+i|e\rangle)/\sqrt{2}
$; after that, a string operation of Eq. (\ref{eq-string}) $%
S^x(t)=\sigma^x_3\sigma^x_4\sigma^x_1\sigma^x_2$ acts on the superposed
state to revolve the $m$ particle around $e$ particle one lap and it results
the appearance of a local phase factor in the state $(|\xi\rangle-i|e%
\rangle)/\sqrt{2}$; finally take a $\sqrt{Z}^{-1}$ operation to annihilate
the anyons to get the state $(|0000\rangle-|1111\rangle)/\sqrt{2}$. Here the
minus sign before the second term arises exactly from the anyons braiding.
To measure changes of the state, we use the state tomography technique to
display all $16\times16$ components for a four-qubit density matrix rather
than using an ancilla to form interferometry in optics. The relative
magnitudes and phase factors between these components can be explicitly
demonstrated out. For comparison, we also measures the state change after a
revolving of $m$ particle in absence of $e$ particle by canceling the $\sqrt{%
Z}$ and subsequent $\sqrt{Z}^{-1}$ operations above.
\begin{figure}[tbph]
\includegraphics[width=0.9\columnwidth]{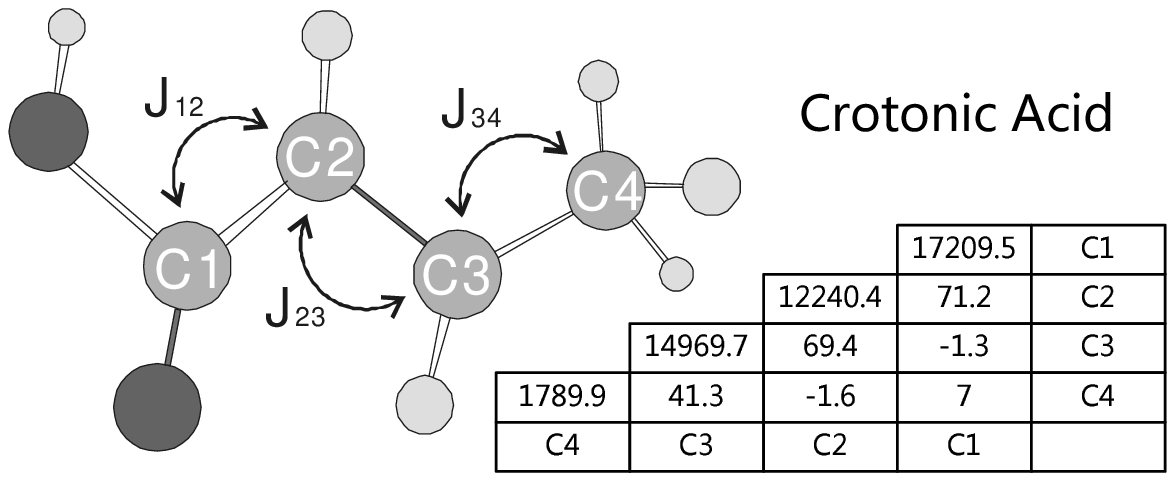} \newline
\caption{Molecular structure and Hamiltonian parameters Crotonic acid. The
chemical shift of each of the carbon nuclei is given by the corresponding
diagonal elements while the coupling strengths are given by the each of the
different off-diagonal elements.}
\label{fig3}
\end{figure}

In our liquid state NMR system, the molecule ($^{13}$C-labeled crotonic)
used for this experiment contains four $^{13}$C spin-1/2 nuclei as qubits,
shown as Fig. \ref{fig3}. The Hamiltonian of the 4-qubit system is (in
angular frequency units) $H_{sys} =\sum_{i=1}^{4}
\omega_{i}I_{z}^{i}+2\pi\sum_{i<j}^{4}J_{ij}I_{z}^{i}I_{z}^{j}$ with the
Larmor angular frequencies of the $i^{th}$ spin $\omega_i$ and spin-spin
coupling constants $J_{12}=71.2$Hz, $J_{13}=-1.3$Hz, $J_{14}=7$Hz, $%
J_{23}=69.4$Hz, $J_{24}=-1.6$Hz, $J_{34}=41.3$Hz. The system Hamiltonian $%
H_{sys}$ exhibits one-dimensional Heisenberg spin chain with neighboring
two-body interactions. Experiments were performed at room temperature using
a standard 400MHz NMR spectrometer (AV-400 Bruker instrument).

The system was first prepared in a pseudo-pure state (PPS) $\rho _{0000}=%
\frac{1-\epsilon }{8}\mathbf{I}+\epsilon |0000\rangle \langle 0000|$, where $%
\epsilon \approx 10^{-5}$ describes the thermal polarization of the system
and $\mathbf{I}$ is a unit matrix, using the method of spatial averaging.
The first part of PPS is the background environment, while the second part
is the effective pure state in use denoted by $\rho_\epsilon$ (not including
$\epsilon$). Then to prepare the GHZ state, a Hadamard gate and three CNOT
gates were needed, shown as the first part of Fig. \ref{fig2}. Finally, a
few single qubit operations were used to do the rotation of $m$ particle
with or without $e$ particle, shown as the other part of Fig. \ref{fig2}.
\begin{figure}[tbp]
\includegraphics[width=0.8\columnwidth]{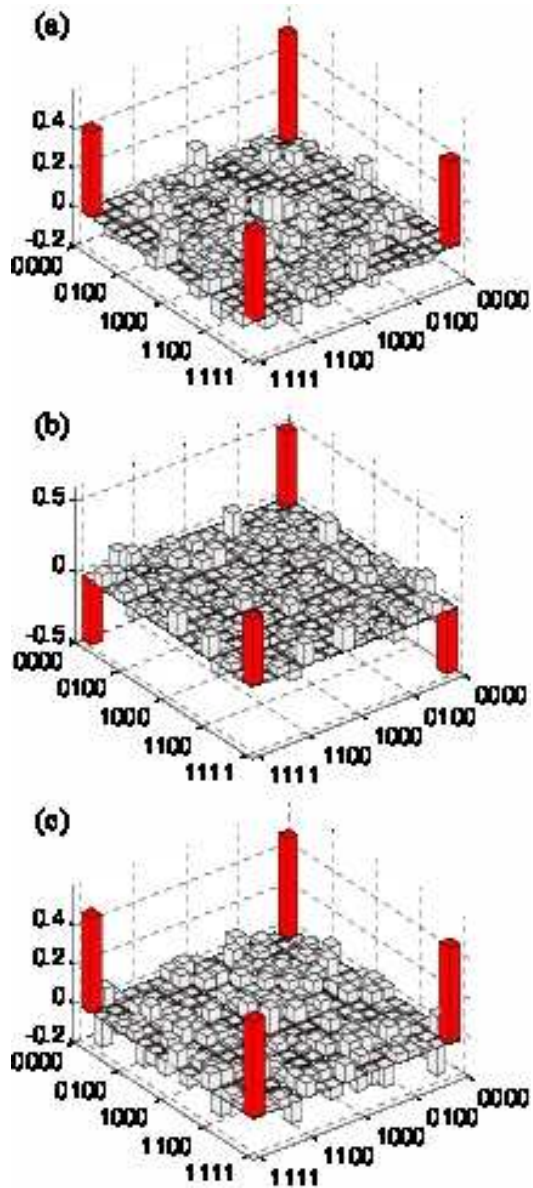}\newline
\caption{(Color online) Experimental results are displayed in the form of
state tomography: (a) the prepared ground state $|\protect\xi\rangle=(1/%
\protect\sqrt{2})(|0000\rangle+|1111\rangle)$; (b) the resultant state due
to the anyons braiding $(1/\protect\sqrt{2})(|0000\rangle-|1111\rangle)$;
(c) the trivial state after a $m$ particle revolving one lap without
braiding $(1/\protect\sqrt{2})(|0000\rangle+|1111\rangle)$, which is used
for comparison with step (b). Theoretically, the height of the red bars
should be $+0.5$ or $-0.5$, and other bars should be $0$.}
\label{fig4}
\end{figure}

In order to improve the quantum coherent control, experimentally every
single qubit gates was created by using robust strongly modulating pulses
(SMP) \cite{16,17,18}. We maximize the gate fidelity of the simulated
propagator to the ideal gate, and we also maximize the effective gate
fidelity by averaging over a weighted distribution of radio frequency (RF)
field strengths, because the RF-control fields are inhomogeneous over the
sample. Theoretically the gate fidelities we calculated for every pulse are
greater than $0.99$, and the pulse lengths range from $200$ to $500$ $\mu s$%
. The quantum circuit of Fig. \ref{fig2} was realized with a sequence of
these local SMPs separated by time intervals of free evolution under the
Hamiltonian. The overall theoretical fidelity of this pulse sequence is
about $0.95$. The density matrices of the spin system after the initial GHZ
state preparation and after braiding were reconstructed using state
tomography, which involves applying 40 readout pulses to obtain coefficients
for the 256 operators comprising a complete operator basis of the four spin-$%
\frac{1}{2}$  $^{13}$C nuclei.

The experimental results are displayed on Fig. \ref{fig4}. The ground state $%
|\xi\rangle$ (i.e., GHZ state) for anyons braiding is prepared at initial
time $t_0$ with fidelity $F(t_0)=\langle\xi|\rho_{\epsilon}(t_0)|\xi%
\rangle=0.93$ and its state tomography is shown on Fig. \ref{fig4} (a).
After braiding, the evolved state $|\xi^{\prime}\rangle$ at time $t_1$
obtain a local phase change with a minus sign instead of a positive sign in
the initial ground state $|\xi\rangle$ and its state tomography on Fig. \ref%
{fig4} has the fidelity $F(t_1)=\langle\xi^{\prime}|\rho_{\epsilon}(t_1)|%
\xi^{\prime}\rangle=0.94$. The different directions of bars reveal the phase
shift, which is just the requirement of the anyonic fractional statistics.
For comparison, we also measure the state $|\xi^{\prime\prime}\rangle$ after
one $m$ particle revolves one lap around empty center at time $t_2$ and it
gives a completely different result from that after braiding, which shows
the topological property of anyons. The state tomography is shown on Fig. %
\ref{fig4} (c) and has the fidelity $F(t_2)=\langle\xi^{\prime\prime}|\rho_{%
\epsilon}(t_2)|\xi^{\prime\prime}\rangle=0.98$. Clearly, the results are in
excellent agreement with the theoretical expectation. The deviation between
the experimental and theoretical values is primarily due to the
inhomogeneity of the radio frequency field and the static magnetic field,
imperfect calibration of radio frequency pulses, and signal decay during the
experiments.

In summary, we are the first to simulate the fractional statistics of the
1/2 abelian anyons by using interacted nuclear spin-1/2 particles in
molecules. The experimental results directly demonstrate the presence of
fractional statistics in a physical system. With the number increase of
controllable qubits to eight, it can realize completely the complete
smallest Kitaev model as discussed previously. This kind of
proof-of-principle demonstration of anyons in small and relatively simple
systems will represent an important step toward the long pursued goal to
demonstrate fractional statistics of quasiparticles in a macroscopic
material. Such abilities, properly extended to large systems, will also be
critical for future implementation of naturally fault-tolerant quantum
computation.

We would like to thank Jian-Wei Pan for inspiring conversations. This work
was supported by the National Natural Science Foundation of China, the CAS,
Ministry of Education of PRC, and the National Fundamental Research Program.
This work was also supported by European Commission under Contact No. 007065
(Marie Curie Fellowship). J.-L. C. acknowledges supports in part by NSF of
China (Grant No. 10575053 and No. 10605013) and Program for New Century
Excellent Talents in University.


\begin{thebibliography}{99}
\bibitem{1982Wilczek} F. Wilczek, Phys. Rev. Lett. \textbf{48}, 1144-1146
(1982).

\bibitem{1982Wilczek2} F. Wilczek, Phys. Rev. Lett. \textbf{49}, 957-959
(1982).

\bibitem{2004Preskill} J. Preskill, \emph{Lectures Notes for Physics 219:
Quantum Computation} Ch. 9 (Online at $<$%
\url{http://www.theory.caltech.edu/people/preskill/ph229/}$>$).

\bibitem{2003Kitaev} A. Yu. Kitaev, Ann. Phys. \textbf{303}, 2-30 (2003).

\bibitem{2006Kitaev} A. Yu. Kitaev, Ann.Phys. \textbf{321}, 2-111 (2006).

\bibitem{1997Preskill} J. Preskill, preprint at $<$%
\url{http://arxiv.org/abs/quant-ph/9712048}$>$ (1997).

\bibitem{2003Mochon} C. Mochon, Phys. Rev. A \textbf{67}, 022315 (2003).

\bibitem{2004Mochon} C. Mochon, Phys. Rev. A \textbf{69}, 032306 (2004).

\bibitem{2000Freedman} M. H. Freedman, M. Larsen, and Z. -H. Wang, preprint
at $<$\url{http://arxiv.org/abs/quant-ph/0001108}$>$ (2000).

\bibitem{2002Freedman} M. H. Freedman, A. Kitaev, and Z. Wang, Comm. Math.
Phys. \textbf{227}, 587-603 (2002).

\bibitem{1982Tsui} D. C. Tsui, H. L. Stormer, and A. C. Gossard, Phys. Rev.
Lett. \textbf{48}, 1559-1562 (1982).

\bibitem{1983Laughlin} R. B. Laughlin, Phys. Rev. Lett. \textbf{50},
1395-1399 (1983).

\bibitem{2005Sarma} S. D. Sarma, M. Freedman, and C. Nayak, Phys. Rev. Lett.
\textbf{94}, 166802 (2005).

\bibitem{2006Stern} A. Stern, and B. I. Halperin, Phys. Rev. Lett. \textbf{96%
}, 016802 (2006).

\bibitem{2006Bonderson} P. Bonderson, A. Kitaev, and K. Shtengel, Phys. Rev.
Lett. \textbf{96}, 016803 (2006).

\bibitem{2007Weeks} C. Weeks, G. Rosenberg, B. Seradjeh, and M. Franz,
Nature Phys. \textbf{3} 796-801 (2007).

\bibitem{2003Duan} L.-M. Duan, E. Demler, and M. D. Lukin, Phys. Rev. Lett.
\textbf{91}, 090402 (2003).

\bibitem{2007Han} Y.-J. Han, R. Raussendorf, and L.-M. Duan, Phys. Rev.
Lett. \textbf{98}, 150404 (2007).

\bibitem{2007Pachos} Pachos et al., preprint at $<$%
\url{http://arxiv.org/abs/0710.0895}$>$ (2007).

\bibitem{2007Lu} C.-Y. Lu et al., preprint at $<$%
\url{http://arxiv.org/abs/0710.0278}$>$ (2007).

\bibitem{2007Jiang} L. Jiang, G. K. Brennen, A. V. Gorshkov, and K.
Hammerer, preprint at $<$\url{http://arxiv.org/abs/0711.1365}$>$ (2007).

\bibitem{2004Vandersypen} L. M. K. Vandersypen, and I. L. Chuang, Rev. Mod.
Phys. \textbf{76}, 1037-1069 (2004).


\bibitem{16} E. Fortunato et al., Chem. Phys. \textbf{116} (17), 7599 (2002).

\bibitem{17} M. A. Pravia et al., J. Chem Phys. \textbf{119}, 9993 (2003).

\bibitem{18} T. S. Mahesh and D. Suter, Phys. Rev. A \textbf{74}, 062312
(2006).
\end{thebibliography}
\end{document}